\documentclass[twocolumn,preprintnumbers,superscriptaddress,nofootinbib,aps,prd,floatfix]{revtex4}
\pdfoutput=1

\usepackage{footmisc,multirow} 
\usepackage{subfigure,tensor,color}
\usepackage{amsmath,slashed,enumerate}
  
\usepackage{graphicx}

\renewcommand{\d}{{\text{d}}}
\newcommand{\ee}{\end{eqnarray*}}

\newcommand{\bee}{\begin{eqnarray}}
\newcommand{\eee}{\end{eqnarray}}
\newcommand{\beeq}{\begin{equation}}
\newcommand{\eeeq}{\end{equation}}

\newcommand{\SU}{\text{SU}}
\newcommand{\SO}{\text{SO}}
\newcommand{\U}{\text{U}}

\newcommand{\Ctop}{C_{\text{top}}}

\newcommand{\hn}{\hat{h}}

\newcommand{\vev}[1]{\langle #1 \rangle} 

\newcommand{\CRRRR}{F^{\tb}_{\textrm{RR}}}
\newcommand{\CRRRRn}{\hat{F}^{\tb}_{\textrm{RR}}}
\newcommand{\CLLLL}{F^{\tb}_{\textrm{LL}}}
\newcommand{\CLLLLn}{\hat{F}^{\tb}_{\textrm{LL}}}
\newcommand{\CtLR}{F_{ \textrm{LR}}^{\tb}}
\newcommand{\CtLRn}{\hat{F}_{ \textrm{LR}}^{\tb}}

\newcommand{\FEWn}{\hat{F}^{\tb}_{\textrm{EW}}}

\newcommand{\tb}{}
\newcommand{\CLR}{C^{\WW}_{\rm LR}}
\newcommand{\CLRn}{\hat{C}^{\WW}_{\rm LR}}
\newcommand{\clrn}{\hat{c}^{\WW}_{\rm LR}}
\newcommand{\WW}{}

\newcommand{\By}{{\cal B}}
\newcommand{\bBy}{\bar{\cal B}}

\newcommand{\al}{\alpha}
\newcommand{\be}{\beta}

\newcommand{\la}{\lambda}
\newcommand{\eps}{\epsilon}

\numberwithin{equation}{section}

\begin{document}

\title{A UV Complete Compositeness Scenario: LHC Constraints Meet The Lattice}

\preprint{CP3-Origins-2017-008 DNRF90}

\begin{abstract} 
  We investigate a recently proposed UV-complete composite Higgs
  scenario in the light of the first LHC runs. The model is based on a
  $\SU(4)$ gauge group with global flavour symmetry breaking
  $\SU(5) \to \SO(5)$, giving rise to pseudo Nambu-Goldstone bosons in
  addition to the Higgs doublet. This includes a real and a complex
  electroweak triplet with exotic electric charges. Including these,
  as well as constraints on other exotic states, we show that LHC
  measurements are not yet sensitive enough to significantly constrain
  the model's low energy constants. The Higgs potential is described
  by two parameters which are on the one hand constrained by the LHC
  measurement of the Higgs mass and Higgs decay channels and on the
  other hand can be computed from correlation functions in the
  UV-complete theory. Hence to exclude the model at least one
  constant needs to be determined and to validate the Higgs potential
  both constants need to be reproduced by the UV-theory. 
   Due to its UV-formulation, a certain number of  low energy constants can be computed from first principle numerical simulations of the theory formulated on a lattice, which can help in establishing the validity of this model. We assess the potential impact of lattice calculations for phenomenological studies, as a preliminary step towards Monte Carlo simulations.
\end{abstract} 

\author{Luigi Del Debbio}
\affiliation{Higgs Centre for Theoretical Physics, School of Physics \& Astronomy, University of Edinburgh, EH9 3FD, UK\\[0.1cm]}

\author{Christoph Englert} 
\affiliation{SUPA, School of Physics and Astronomy, University of Glasgow, Glasgow G12 8QQ, UK\\[0.1cm]}

\author{Roman Zwicky} 
\affiliation{Higgs Centre for Theoretical Physics, School of Physics \& Astronomy, University of Edinburgh, EH9 3FD, UK\\[0.1cm]}

\maketitle


\tableofcontents

\section{Introduction}
\label{sec:intro}

The discovery of a Standard Model (SM) like Higgs boson has so far not
revealed concrete hints towards an understanding of the electroweak
symmetry breaking (EWSB).  The concept of Higgs naturalness stands
questioned in many established BSM scenarios such as supersymmetry but
also in theories of Higgs compositeness. It is conceivable that future
LHC runs, exploring higher energy scales with large statistics, will
improve the situation. Due to the non-perturbative nature of the
composite Higgs models, their phenomenological investigations are
typically informed by means of effective theories, in a way that is
completely analogous to the description of the low energy dynamics of
QCD by chiral perturbation theory. Although these methods have been
very successful in understanding the low energy properties of QCD, the
ultimate goal is obviously to analyse the phenomenological properties
of a composite Higgs scenario by investigating concrete UV-complete
candidate theories using non-perturbative techniques to gain a more
complete picture of their dynamics from first principles.

The minimal composite Higgs model
(MCHM)~\cite{Contino:2003ve,Agashe:2004rs,Contino:2006qr} based on
global symmetry breaking pattern $\SO(5)\to \SO(4)$ is the prototype of
composite Higgs model.  The four arising Nambu-Goldstone bosons (NGBs)
transform as a bi-doublet under $\SU(2)\times \SU(2) \simeq \SO(4)$ and
can therefore be identified with the Higgs doublet in the SM. Breaking
the global symmetry by gauging the weak interaction
$\SU(2)_L\times \U(1)_Y \subset \SO(4)$ in the presence of heavy
composite fermions induces a Higgs potential. Whether or not the
potential triggers EWSB can only be investigated for definite in a UV
complete scenario.  The scale of the composite sector is parametrised
by $f$ and its value compared to $v$, $\xi \equiv v^2/f^2 $
($v\simeq 246~\text{GeV}$ and $\xi \leq 0.12$, e.g.~\cite{Aad:2015pla}), is a
measure of the misalignment of the new strong sector and the Higgs
sector vacuum.  Low energy scenarios based on this symmetry breaking
pattern have been scrutinised in the literature in
detail~\cite{Espinosa:2010vn,Azatov:2012qz,Azatov:2012bz,Espinosa:2012im},
however, no UV complete realisation of this minimal scenario has been
established so far (see~e.g. Ref.~\cite{Elander:2013jqa} for related
work in the holographic context).

Recently Ferretti, in Ref.~\cite{Ferretti:2014qta}, proposed a UV
complete model based on a $\SU(4)$ (hypercolor) gauge symmetry with a
flavour structure motivated by partial compositeness, gauge anomaly
cancellation and asymptotic freedom \cite{FK13}.  Earlier UV-complete realisations of the (partial) composite Higgs scenarios
  are based on embedding effective four fermion operators into gauge
  theory~\cite{Barnard:2013zea} or non-negligible irrelevant
  operators of SM and composite fermions~\cite{Vecchi:2015fma}.
 
This model is distinct, and non-minimal, as compared to MCHM4/5 in
that the flavour structure predicts a number of extra pseudo NGBs
(PNGB).  In this work we reflect on the potential impact of lattice
studies on the Higgs sector (e.g. Higgs potential, mass spectrum,
\dots) and investigate the LHC phenomenology of the exotic extra
PNGBs. The combined analyses of Higgs measurements and LHC constraints
on exotics allows us to identify a region of parameter space of the
model, which can be cross checked against lattice calculations. This
provides an important guideline for future efforts to construct,
modify, simulate and validate UV-complete models of Higgs
compositeness. 
Pioneering lattice
studies~\cite{DeGrand:2016mxr,DeGrand:2016htl} have shown that these
simulations are computationally demanding, therefore strengthening the
case for a detailed understanding of the lattice measurements that
will be relevant for phenomenology.

This work is organised as follows: In Sec.~\ref{sec:model} we briefly
summarise the model of~\cite{Ferretti:2014qta} to make this work
self-contained.  The relevant low energy constants (LECs) which can be computed 
on the lattice are discussed and identified.
Subsequently, in Sec.~\ref{sec:results} we study the
model with available LHC Higgs measurements, for which preliminary results have been presented
in \cite{DEZproc}, and  include
constraints from searches for predicted exotic states, which have so far
not been discussed in the literature.
We summarise and conclude in Sec.~\ref{sec:summary}.

\section{Ferretti's model}
\label{sec:model}

Ferretti's model~\cite{Ferretti:2014qta} is a gauge theory with
hypercolour gauge group $G_{\text{HC}}=\SU(4)$ with 5 massless Weyl
fermions transforming in the two-index antisymmetric representation of
$G_{\text{HC}}$, and 3 massless Dirac fermions in the fundamental
representation of color. Using Weyl fermions, we denote these fermions
$\psi,\chi,\tilde\chi$ respectively, with $\psi \in \bf{6}$ and
$\chi \in \bf{4}, \tilde \chi \in \bar{\bf{4}}$ under $G_{\text{HC}}$.
The theory has a global symmetry group 
\begin{equation}
\label{eq:break}
G_F = \SU(5)\times \SU(3) \times \SU(3)' \times \U(1)_X\times \U(1)' \, .
\end{equation}
The strong dynamics of $G_{\text{HC}}$ is expected to break the global
flavour symmetries $\SU(5)\to \SO(5)$ and
$\SU(3)\times \SU(3)'\to \SU(3)_c$, as well as  $\U(1)_X$.\footnote{$\SU(5)\to \SO(5)$ has also been considered in the littlest Higgs model~\cite{ArkaniHamed:2002qy}. See also \cite{Terazawa:1976xx,Terazawa:1979pj} for early discussions of $\SU(4)$ gauge symmetries in strongly interacting theories.}
The maximally attractive channel hypothesis~\cite{Raby:1979my} 
suggests $\SU(5)\to \SO(5)$ to occur at a higher scale than
$\SU(3)\times \SU(3)'\to \SU(3)_c$. This leads to a low-energy effective
theory based on the global symmetry breaking pattern
\begin{alignat}{5}
  G_F/H_F &= {\SU(5)\times \SU(3) \times \SU(3)' \times \U(1)_X\times \U(1)' \over \SO(5) \times \SU(3) \times \U(1)_X}\nonumber\\
  \label{eq:symbreak}
  &= {\SU(5)\over \SO(5)} \times {\SU(3) \times \SU(3)'\over \SU(3)} \times
  \U(1)'\,.
\end{alignat}
Since $\SO(5)\supset \SO(4)\simeq \SU(2)\times \SU(2)$, the unbroken
global symmetry group $H_F$ contains the custodial subgroup
\begin{equation}
  H_c=\SU(2)_L\times \SU(2)_R \, .
\end{equation}
Following the standard paradigm of composite Higgs scenarios, the SM
subgroup
$ G_{\text{SM}} \equiv \SU(3)_c\times \SU(2)_L \times \U(1)_Y \subset H_F
$ is weakly gauged and the hypercharge is a linear combination of
$\SU(2)_R$ and $\U(1)_X$, $Y=T_R^3 + X$. Weakly gauging a subgroup and
heavy quark mass generation through partial
compositeness~\cite{Kaplan:1991dc,Contino:2006nn} amount to explicit
violation of $G_F$, and the analysis of the one-loop effective
action~\cite{Ferretti:2014qta} shows that this indeed gives rise to
NGB misalignment and EWSB
$\SU(2)_L\times \U(1)_Y\to \U(1)_{\text{QED}}$, in a way that is
completely analogous to the minimal effective
realisations~\cite{Agashe:2004rs, Agashe:2006at,Contino:2006qr}. The
difference between the MCHM4/5 scenario of~\cite{Contino:2006qr} is the
prediction of 14 NGBs from the $\SU(5)\to \SO(5)$ breaking. 
The NGBs fields are denoted by $\Pi$, classified  according
to their  $\SU(2)_L\times \U(1)_Y$ quantum numbers 
$\Pi= (\eta,H,\Phi_0,\Phi) \in {\mathbf{1}}_0 + {\mathbf{2}}_{\pm 1/2} +
{\mathbf{3}}_0 + {\mathbf{3}}_{\pm 1}$ and 
the ${\bf{2}}_{\pm1/2}$ is identified as the SM Higgs doublet. 
In this work we investigate the phenomenology of the triplet states 
but ignore the NGB-singlets $\eta$ mentioned above and $\eta'$ due to $\U(1)'$-breaking whose 
phenomenology has been scrutinised in ~\cite{Ferretti:2016upr,Belyaev:2016ftv}.

This extended scalar sector reveals parallels with the so-called
Georgi-Machacek
model~\cite{Georgi:1985nv,Chanowitz:1985ug,Gunion:1989ci} (for recent
phenomenological investigations see
also~\cite{Englert:2013zpa,Bambhaniya:2015wna,Logan:2015xpa,Degrande:2015xnm,Hartling:2014aga}),
which also predicts the appearance of a real as well as a complex
$\SU(2)_L$ triplet in the scalar sector. 
Whether or not these extra states contribute to the breaking of electroweak
symmetry, as in the Georgi-Machacek
model, is an interesting open question to be addressed in future research~(see e.g.~\cite{Golterman:2017vdj} for similar considerations). We will follow Ferretti's original ansatz where the triplet states do not contribute to electroweak symmetry breaking.
The construction of the
low-energy effective theory follows the approach pioneered by Callen,
Coleman, Wess and Zumino (CCWZ)
~\cite{Coleman:1969sm,Callan:1969sn}. Denoting the $\SU(5)/\SO(5)$
generators by $T^{\hat{A}}$, a non-linear sigma field is introduced 
\begin{equation}
\label{eq:f}
	\Sigma(x) = \exp\left( {i \Pi \over f} \right)\,,~\Pi = \phi^{\hat{A}}(x) T^{\hat{A}}\,,
\end{equation}
 transforming non-linearly $\Sigma \to \mathrm{g} \, \Sigma  \, \mathrm{h}$ 
 since  $\mathrm{h} \in \SO(5)$ is $\Pi$- and $\mathrm{g}$-dependent.
 The quantity  $f \equiv f_{\SU(5)/\SO(5)}$ is the $\Pi$
decay constant which can be thought of as setting the scale of the
hypercolour gauge theory.  Since $\SU(5)/\SO(5)$ is a symmetric space,
 the CCWZ kinetic term, governing the interactions with the gauge bosons, is simplified to 
\begin{equation}
\label{eq:16}
	{\cal{L}}\supset {f^2\over 16} \text{Tr}\left( D_\mu U D^\mu U^\dagger \right) \,,
\end{equation}
where  $U=\Sigma\Sigma^T=\exp(2i\Pi/f)$
transforms linearly 
$U\mapsto \mathrm{g}\,U \mathrm{g}^T$ under $\mathrm{g}\in
\SU(5)$. The covariant derivative is given by
\begin{equation}
	D_\mu U= \partial_\mu U - ig W^A_\mu[T^A_L,U] - ig' B_\mu [T^3_R,U] \,,
\end{equation}
as all NGBs have zero $\U(1)_X$ charge.  With the convention
${\text{tr}}[T^A T^B]= \delta^{AB}/2$ and
$ \Pi\supset H^+ T^+ = H^+\sqrt{2}(T^{18}-iT^{15})$, Eq.~\eqref{eq:16}
leads to canonically normalised kinetic terms.

Expanding this Lagrangian we find the standard MCHM4/5 coupling
modifications of the physical Higgs boson to the massive electroweak
gauge bosons rescaled by $\sqrt{1-\xi}$, where $\xi \equiv v^2/f^2$, while the remaining PNGB interactions
are completely determined by their $\SU(2)_L$ quantum numbers.

Heavy third family quark masses are included through partial
compositeness~\cite{Kaplan:1991dc,Contino:2006nn}, i.e.~mixing effects
with vector-like hyperbaryons of the strongly interacting sector.  The
relevant terms originating from an extended HC (EHC) sector are
\begin{multline}
\label{eq:massint}
	-{\cal{L}} \supset {M} \bar\Psi \Psi + \lambda_q f \bar{\hat{q}}_L\Sigma \Psi_R + \lambda_t f \bar{\hat{t}}_R \Sigma^\ast\Psi_L \\ +
	\sqrt{2} \mu_b \text{Tr}( \bar{\hat{q}}_L^3 U \hat{d}^3_R ) + \text{h.c.}
\end{multline}
where we introduced the field $\Psi$ to represent the composite
fermion in the effective theory, transforming under a $\bf{5}$ of
$\SO(5)$ and a $\bf{3}$ of $\SU(3)_c$, and $\hat{q}_L\supset (t_L,b_L)$,
and $\hat{t}_R\supset t_R$ are $\SO(5)$-spurionic embeddings of the
third generation quarks.  The field $\Psi$ can be written in terms of
its components that have definite quantum numbers under the standard
model gauge group $\SU(3)_c\times \SU(2) \times \U(1)$:
\begin{alignat}{5}
  \label{eq:toppartner}
    \Psi      &= \frac{1}{\sqrt{2}} [iB - iX, B+X, iT+iY, -T+Y,\sqrt{2}iR]\,,
\end{alignat}
where the quantum numbers are
$(T,B) \in (\mathbf{3},\mathbf{2})_{1/6}$,
$R \in (\mathbf{3},\mathbf{1})_{2/3}$, and
$(X,Y) \in (\mathbf{3},\mathbf{2})_{7/6}$. Expanding this Lagrangian
yields a mass matrix in the top partner space $(t,T,Y,R)$:
\begin{multline}
  \label{eq:topmass}
  \hat{{\cal{M}}}_T=\\\left( \begin{matrix} 0  & {\lambda_q\over 2}  (1+c_h)  & {\lambda_q\over 2}  (1-c_h) & {\lambda_q\over \sqrt{2} }  s_h\\
      {\lambda_t\over\sqrt{2}}   s_h & \hat{M} & 0 & 0 \\
      -{\lambda_t\over\sqrt{2}}   s_h	& 0 & \hat{M} & 0 \\
      \lambda_t  c_h & 0 & 0 & \hat{M}
    \end{matrix}\right)\, ,
\end{multline}
and an analogous matrix in the bottom partner space $(b,B)$: 
\begin{equation}
  \label{eq:bottmass}
  \hat{{\cal{M}}}_B=\\\left( \begin{matrix} \hat{\mu}_bs_hc_h  & \lambda_q  \\ 0 & \hat{M} 
 \end{matrix}\right) \, ,
\end{equation}
where hatted quantities, e.g.~$\hat{M} \equiv M/f$, are
made dimensionless by dividing by the appropriate power of $f$.
In the expressions above $c_h \equiv \cos(\hat{h})$ and
$s_h \equiv \sin(\hat{h})$, where $h$ is the physical Higgs in the
unitary gauge.
Bi-unitary transformations yield the physical top and bottom partner
mass spectrum as well as their (non-diagonal) interactions with the
Higgs after expanding $s_h,c_h$.  Note that the $X$-particle and the
Higgs $h$ do not interact at the tree-level.  To lowest order in $v$
the top mass ${\cal{O}}(v^0)$ and bottom mass ${\cal{O}}(v)$ are given by
\begin{equation}
  \label{eq:topmassl}
  m_t\simeq \frac{\sqrt{2}  \lambda_q \lambda_t  }{\sqrt{ \hat{M}^2+\lambda_q^2} \sqrt{\hat{M}^2+\lambda_t^2}} M
\end{equation}
and
\begin{equation}
  \label{eq:botmassl}
  m_b\simeq \frac{   \hat{M} \hat{\mu}_b }{ \sqrt{\hat{M}^2+ \lambda_q^2}} v\,,
\end{equation}
where $v = \sin ( \vev{\hat{h}})$ has been used in the last equation.
It is seen from Eq.~\eqref{eq:botmassl} that $\hat{\mu}_b$ essentially acts like a Yukawa coupling 
for the $b$-quark as in the SM.
Eq.~\eqref{eq:topmassl} is inverted to $\lambda_q=\lambda_q(m_t)$
for the scan for which we use $m_t\simeq 173~\text{GeV}$.  We use a
similar strategy to invert Eq.~\eqref{eq:botmassl}
$\mu_b=\mu_b(m_b,\lambda_q)$ with $m_b\simeq 4.7~\text{GeV}$ as an
input.  Furthermore, we will require $M > 1.5~\text{TeV}$ (see below)
and leave $f$ as a free parameter.

The SM-like Higgs boson phenomenology is identical to MCHM4/5 but
includes the previously mentioned exotically charged NGBs.  The masses
of the NGBs are radiatively induced, in analogy to the $\pi^\pm -\pi^0$
mass difference in the SM due to electromagnetic interaction.  The
leading order expression assumes the form~\cite{Golterman:2015zwa}\footnote{
A further contribution to $\Phi^2$ from the integrating out the third generation quarks. 
C.f. the Higgs potential section for further remarks.}
\begin{equation}
  \label{eq:potential}
  V= f^2 \CLRn    \left( (3g^2+g'^2) \left( 2H^\dagger H + {16\over 3} \Phi^\dagger \Phi \right) + 8 g^2 \Phi_0^\dagger \Phi_0\right)\,,
\end{equation} 
where $3 g^2 + g'^2 \simeq 1.31$ and $g^2 \simeq 0.40$ and
\begin{equation}
  \label{eq:CLR}
  \CLR= {3\over 16 \pi^2} \int_{0}^\infty \hbox{d}q^2\, q^2 \Pi^{33}_{LR}(q^2) \,,
\end{equation}
is an integral over the $\SU(2)_L\times \SU(2)_R$-correlator
\begin{equation}
  i \int \d^4 x\, e^{iq \cdot x} \,  \,\langle  T J^{\mu \,a}_{L}(x) \,  J^{\mu\,b}_{R}(0) \rangle  = 
    \Pi^{ab}_{LR}(q^2)\, P^{\mu \nu}  \, .
\end{equation}
Above $P^{\mu \nu} = ( q^2 g^{\mu\nu} -q^\mu q^\nu )$,
$g^{\mu\nu} = {\text{diag}}(1,-1,-1,-1)$ and the chiral currents are
in the adjoint flavour representation
$2 J^{\mu\,a}_{L,R}=\bar \psi \gamma^\mu (1\mp \gamma^5) T^a \psi$.
This current has the right quantum numbers to excite the NGBs and
therefore $\lim_{q^2 \to 0} q^2 \Pi^{33}_{LR}(q^2)=f^2$ as the lowest
term in a $q^2$ expansion, which underpins Eq.~\eqref{eq:potential}.
In the next section we will consider further corrections to the Higgs potential 
for which LHC  constraints furnish a value for $\CLRn$. 
The latter  gives a lower bound on the triplet masses $\Phi$ and $\Phi_0$.
Further low energy exotic states
include an $\SU(3)_c$ octet hyper-pion, whose mass is estimated to be
in the multi-TeV regime~\cite{Ferrera:2011bk} and has  been investigated
phenomenologically in~\cite{Belyaev:2016ftv}.

\subsection*{Lessons from the Lattice and the LHC}
\label{sec:lattice}
Several LECs are accessible by first principle computations,
e.g.~lattice Monte Carlo simulations, of the UV complete theory.  As
previously mentioned one might think of $f$, the $\Pi$ decay constant
in Eq.~\eqref{eq:f}, as setting the scale of the $\SU(4)$-hypercolour
theory.  In increasing order of complexity LECs of interest are the
spectrum of the lowest lying state in a given channel (including the
composite baryon mass $\hat{M}$), the quark condensates
$\vev{\Psi\Psi}$ and $\vev{\bar \chi \chi}$ with associated decay
constants $f$ and $f'$, and the Higgs potential parameters {\color{red} $\CLRn$, 
$\FEWn$ and $\CLLLLn$}resulting from non-trivial correlation functions. Preliminary
lattice investigations have already
started~\cite{DeGrand:2016mxr,DeGrand:2016htl}, highlighting the
subtleties involved in simulating models with fermions in multiple
representations of the gauge groups. The results in this section
should help in identifying the lattice measurements that are likely to
have a significant phenomenological impact.

\subsubsection*{Higgs Potential}
\label{sec:higgs-potential}

As discussed above, the Higgs particle is one of the NGBs of the UV
complete theory. In the hypercolour theory in isolation, no potential
is generated for the NGBs; hence the Higgs potential can only arise
from interactions with the SM sector. In particular there are two
contributions to the one-loop effective potential: the first one is
due to the coupling to the weak gauge bosons (cf. Eq.~\eqref{eq:potential})
and the second one to the
coupling to the top and the composite fermions.  Using the standard
composite Higgs potential parametrisation
\begin{equation}
  \label{eq:HiggsPotAgain}
  \hat{V}(\hn  ) =  \alpha \cos (2 \hn ) - \beta
  \sin^2(2 \hn )\, ,
\end{equation}
the dimensionless parameters $\al$ and $\be$ are given by 
\begin{alignat}{2} 
\label{eq:ab-corr}
& \al   
   \nonumber &=&   \frac{1}{2} \CLLLLn - \clrn   \\[0.1cm]
&  \be    &=& 
 \frac{1}{2} \FEWn - \frac{1}{4} \CLLLLn   \;.
\end{alignat}
The quantities 
$\clrn$,  $\CLLLLn$ and  $ \FEWn$ are related to correlation functions of the UV theory.
The quantity 
$\clrn \equiv \frac{1}{2}  ( 3 g^2 + g'^2) \CLRn $ is the previously defined 
$2$-point function Eq.~\eqref{eq:CLR} whose evaluation on the lattice is a routine matter. 
The quantities $\CLLLLn$ and  $ \FEWn  \equiv \CtLRn - 2 \CRRRRn$ are related to 
$4$-point functions as defined in appendix \ref{app:4}. Their evaluation is a more complex task for lattice Monte Carlo simulations.\footnote{Note, Eq.~\eqref{eq:HiggsPotAgain} includes radiative corrections of the type
discussed in~\cite{Foadi:2012bb} in a more systematic way.}  

We can now analyse the potential in terms of $\al$ and $\be$, imposing the Higgs mass
and direct search constraints, and then discuss the relation of the Higgs sector with the two triplet PNGBs. 
Up to a constant the potential Eq.~\eqref{eq:HiggsPotAgain} can be written as 
\begin{equation}
\label{eq:xipot}
\hat{V}(\hat{h}) = 4 \be ( \sin^2 ( \hat{h})  -  \xi  ) ^2 \,,  \quad  
\end{equation}
with 
\begin{equation}
\label{eq:xi}
\xi  \equiv   \frac{v^2}{f^2} = \sin^2 ( \vev{\hat{h}})   =   \frac{\al + 2 \be}{4 \be} \,.
\end{equation}
The important condition for EWSB, then reads
\begin{equation}
  \alpha+2\beta = \FEWn - \clrn  >0\,.
\end{equation}
Hence, the sign of $\beta$, and its magnitude compared to $\alpha$, are
the first constraints that the UV complete theory needs to satisfy.

\begin{figure}[!t]
  \centering
  \includegraphics[width=.40\textwidth]{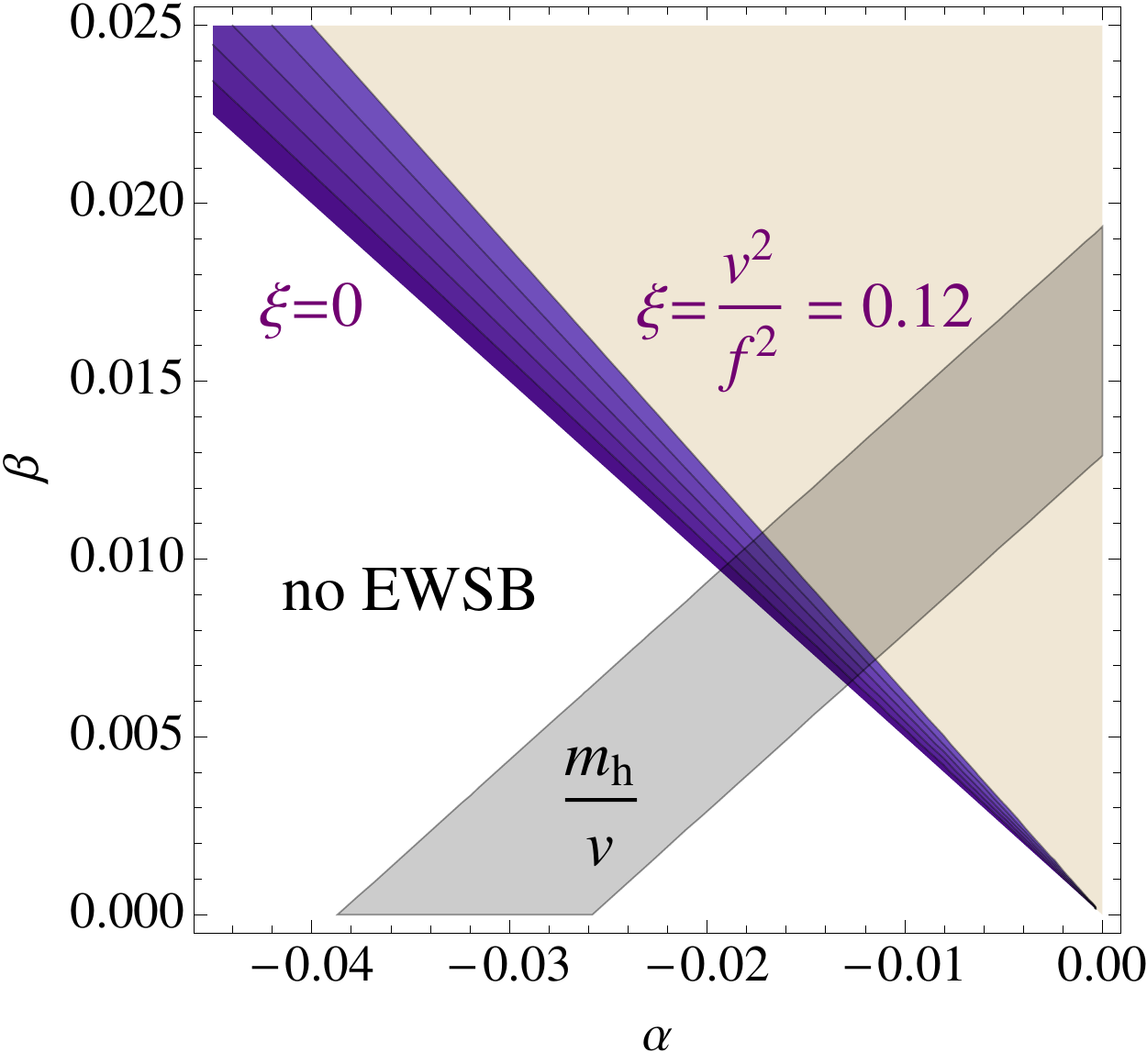}  
  \caption{Contour plot for
    $\xi  = (\al + 2 \be)/(4 \be) $, Eq.~\eqref{eq:xi}.  In the
    white region no EWSB occurs and the purple level curves are values
    of $\xi $ ranging from $0$ to $0.12$ where the latter value is a representative constraint  taken from Ref.~\cite{Aad:2015pla}. An additional constraint comes  from the Higgs mass
    $m_h^2/v^2  = 8( 2 \be- \al) \simeq 0.258 $, Eq.~\eqref{eq:power}, for which we have allowed  generous 
    $20\%$ radiative corrections. The intersection of the purple and grey region is the physically allowed parameter space of the model that has to be satisfied by the UV theory.}
  \label{fig:ABPlot}
\end{figure}

The $\al$-$\be$ parameter space is shown in Fig.~\ref{fig:ABPlot}
with phenomenologically acceptable values of $\xi \in [0,0.12]$ shown
in purple. The Higgs mass is related to the second derivative of the potential 
\begin{multline}	
  \label{eq:HiggsMass}
  \hat{m}_h^2 = \hat{V}''( \vev{\hat{h}})  =  32 \be \xi ( 1- \xi) 
      =  8 \be -2 \al^2/\be  \, ,
\end{multline}
and gives a second constraint, cf. Fig.~\ref{fig:ABPlot}, in the $\al$-$\be$ plane by combining Eqs.~\eqref{eq:xi}
and~\eqref{eq:HiggsMass} 
\begin{equation}
\label{eq:power}
  \frac{m_h^2}{v^2}  =  32 \be  ( 1- \xi) 
  = 8 ( 2 \be - \al) \simeq 0.258  \,.
\end{equation}
From Fig.~\ref{fig:ABPlot}, $0.012 <  -\al < 0.02 $ and $0.06  < \be < 0.11$ can be inferred. Note that this range 
mainly depends on unknown radiative corrections to the Higgs mass. 

Further observables are the two triplet $\Phi_0 \in {\mathbf{3}}_0 $ and $\Phi \in {\mathbf{3}}_1$ (for $\SU(2)_L \times U(1)_Y$) PNGB masses. At leading order 
the mass of $\Phi_0$ is determined by integrating out the gauge bosons Eq.~\eqref{eq:potential}; the charged triplet receives a contribution from integrating out the third generation 
through the $4$-point function $ \CLLLLn$ defined in appendix \ref{app:4},
\begin{alignat}{3}
\label{eq:split}
&  \hat{m}_{\Phi_0}^2 &\;=\;& 16 g^2 \CLRn \;,  &\;\simeq\;&  0.36^2   \;, \\
&  \hat{m}_{\Phi}^2 &\;=\;& 16 (g^2 + \frac{g'^2}{3}) \CLRn + 8 \CLLLLn 
&\;\simeq\;& 0.34^2 + 
8 \CLLLLn \,. \nonumber
\end{alignat}
The triplet masses are equal in the limit where the hypercharge disappears 
$g'^2 \to 0$ and the EHC-coupling $\la_1 \to 0$ (cf.  appendix \ref{app:4}.)  
In the limit  $ \CLLLLn \to 0$   the mass difference of the charged to neutral is positive,
$m_{\Phi} -  m_{\Phi_0} \geq 0 $, as for the pions in the SM \cite{Witten:1983ut}.

From the LHC bound $\xi = v^2/f^2 < 0.12$ it follows that $f \gtrsim 5.7 m_h$ and therfore
\begin{equation}
\label{eq:trip-bounds}
m_{\Phi_0} > 1.97 m_h \,. 
\end{equation}
A lattice determination of  $\CLRn$ ($0.012 <  -\al < 0.02 $ and
$0.06  < \be < 0.11$ for our example) together with Eq.~\eqref{eq:ab-corr} allows us to set an upper and lower bound 
on $\FEWn$ and $\CLLLLn$. The latter can then be tensioned against triplet mass 
$m_{\Phi}$ \eqref{eq:split} and potential lattice determinations of 
$\FEWn$ and $\CLLLLn$.

 In summary the Higgs potential is parametrised by the two constants
$\al$ and $\be$, Eq.~\eqref{eq:HiggsPotAgain}, which are experimentally
constrained by $m_h/v$, $v/f$ and the requirement of EWSB.  On the
other hand $\al$ and $\be$ can be determined from well-defined
correlation function of the UV hypercolor theory, Eq.~\eqref{eq:ab-corr}.
Hence the determination of either $\al$ or $\be$ alone can exclude the
model. As previously mentioned and discussed further below
the computation of $\CLR$ is standard on the lattice
whereas the evaluation of $\FEWn$ and $\CLLLLn$ is far from clear. 
Since no linear combination of $\alpha$ and $\beta$ is independent of 
the $4$-functions $\FEWn$ and $\CLLLLn$ it is therefore not possible for 
the lattice alone to exclude or validate the model but one needs to take
into account further phenomenological consideration discussed in the previous paragraph.

The quantity $\CLR$ has been computed
recently in \cite{lattice-al} for an $\SU(4)$ gauge theory in the
quenched approximation with fermion $N_{f({\bf 6})}= 4$ in the
two-index antisymmetric representation for which we extract a value of
$\clrn \simeq 0.08$. This can only be considered a rough benchmark value since 
Ferretti's model  ($N_{f({\bf 6})}= 5$ and $N_{f({\bf 4})} = N_{f({\bf \bar{4}})}= 3$) differs 
from theirs  ($N_{f({\bf 6})}= 4$ and 
$N_{f({\bf 4})}=  N_{f({\bf \bar{4}})} = 0$)  \cite{lattice-al}).
The feasibility of computing $\CLR$ on the lattice depends on how
quickly the $\CLR$-integral Eq.~\eqref{eq:CLR} saturates in
$q^2$. One can envision to approach this by either computing
$\Pi_{\textrm{LR}}(q^2) $ for low values of $q^2$ observing
convergence or saturate the correlator in the hadronic picture with
the $J^{\textrm{PC}} = 1^{--}$ and $1^{+-}$ $\SU(2)_L$-triplet states
following the idea of the original Weinberg sum rules \cite{Weinberg:1967kj}.  The
computation of $\FEWn$ and $\CLLLLn$ is a formidable task which becomes more
feasible when integrating out the top quarks at ${\cal{O}}(\al_s^0)$ further
neglecting the top quark mass. Even in this case the question of
convergence of the $4$-point correlation function \cite{DEZ17} is far
from trivial since for instance the EHC sector has not been specified \cite{Ferretti:2014qta}.
 In  AdS/CFT  inspired 5D-models the Higgs potential is found to be insensitive 
to the UV-completion \cite{Agashe:2004rs,Contino:2006qr}. It should be noted that this 
scenario automatically assumes a large $N_c$ limit.

\subsubsection*{Quark Condensates and Goldstone Boson Decay Constants}
\label{sec:constr-param-scan}

Quark condensates are related to the zero eigenvalue density of the
Dirac operator via the Banks-Casher relation and can therefore be
studied on the lattice.  Whereas the order parameter for SSB of the
flavour symmetries $\SU(5) \to \SO(5)$ and
$\SU(3)\times \SU(3)' \to \SU(3)_c$ are the corresponding decay
constants, a non-zero or zero value of the corresponding
fermion-condensates $\vev{\Psi\Psi}$ and $\vev{\bar \chi \chi}$
reveals further information about the mechanism of SSB.  Furthermore,
this permits the possibility to check the Gell-Mann Oakes Renner
relation $f'^2 m_\pi^2 = 2 m_\chi \vev{\bar \chi \chi} $ since lattice
simulation are performed at finite quark mass in practice.  A further
possibility is to test the successful Pagels-Stokar relation
\cite{PS79} based on a fermion self energy of the form
$\Sigma(q^2) \to \Lambda^3_{\text{ hadron}} /q^2$ for $q^2 \to \infty$
which can also be motivated from the operator product expansion
\cite{P76}.

\section{Parameter regions after LHC measurements}
\label{sec:results}

The UV complete model in this paper comes with a $\SU(5)$ flavour
symmetry in the Higgs sector which leads to a number of additional
PNGBs, as compared to the MCHM4/5, with exotic charge numbers.  More
precisely, the model predicts the previously mentioned ${\bf{3}}_{0}$
and ${\bf{3}}_{1}$ states, $\Phi_0 \supset ( \phi_0^-,\phi_0^0,\phi_0^+)$ and
$\Phi_+ \supset ( \phi_+^-, \phi_+^0 ,\phi_+^+)$, where the sum of the
$\pm$ and $0$ indices indicate the electric charges $Q = Y + T_L^3$. Additional
exotic particles are the top and bottom partner of the hypercolor
theory.  The NGBs acquire a mass from integrating out the weak gauge bosons
with masses given by Eq.~\eqref{eq:potential} from where 
the ratios to the Higgs masses 

The masses of the weak PNGBs are indirectly constrained by the LHC data through 
$\CLR$, Eq.~\eqref{eq:potential} which leads to  $m_{\Phi},m_{\Phi_0} \gtrsim 2 m_h $ at leading order in the effective theory.
Note that at leading order in the effective theory there is no mixing between the two triplet states.
We treat $m_{\Psi}$ and $m_{\Psi_0}$  as
free parameters in our scan in the range $m>200~\text{GeV}>m_h$. We limit our study to
$m<1~\text{TeV}$ due to a vanishing LHC sensitivity.

 We will focus in this work on direct constraints, but a few comments regarding constraints from electroweak precision observables are in order. As the assignment of top partner quantum numbers is analogous to that of MCHM5, the right-left symmetry required to avoid tension with non-oblique $Zb\bar b$ measurements~\cite{Agashe:2006at} is also present in this model and the discussion of gauge and fermion contributions to the oblique parameters follows Refs.~\cite{Agashe:2005dk,Agashe:2006at,Lodone:2008yy,Gillioz:2008hs,Anastasiou:2009rv}. In particular, the numerical analysis of Ref.~\cite{Gillioz:2012se} suggests that electroweak precision constraints can be satisfied over a broad range of values of $\xi$. We can expect the impact of the additional scalars to be further suppressed compared to the top partners. Since they  do not contribute to electroweak symmetry breaking, their loop contribution to the $2$-point electroweak bosons' polarisation functions is entirely due to their gauge interactions. Hence  the constraints from oblique corrections do only constrain the mass splittings between the custodial quintet, triplet and singlets. Due to Eq.~\eqref{eq:trip-bounds} we can expect these contributions to be small.

\subsection*{Constraints from coloured Exotica}

The LHC analysis programme that targets the phenomenology of the
fermionic partners of Eq.~\eqref{eq:toppartner} is well-developed
across a range of final states (see e.g.~\cite{Aad:2012uu}
or~\cite{Sirunyan:2016ipo}). A comprehensive interpretation of
searches for exotic top partner spectra as detailed above has been
performed recently in Ref.~\cite{Matsedonskyi:2015dns}. In particular,
searches for the baryon $X$, with exotic charge $5/3$, set constraints
on the vector-like mass $M\gtrsim 1.5~\text{TeV}$. We include this
constraint to our scan directly.

\begin{figure*}[!t]
  \subfigure[~$\gamma\gamma$ signal
  strength.]{\includegraphics[height=5.8cm]{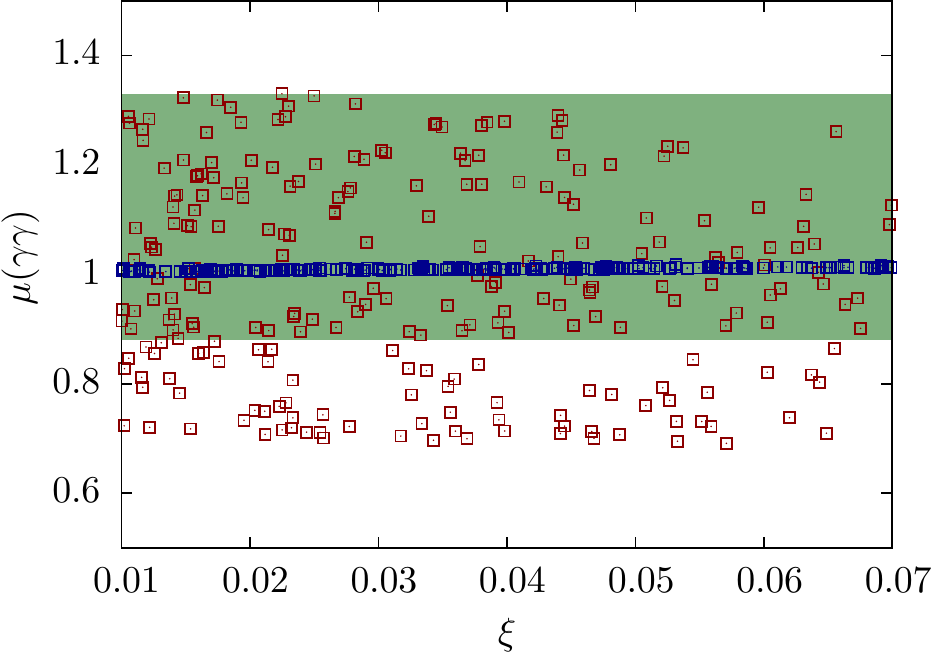}}\hfill
  \subfigure[~$ZZ$ signal strength.]{\includegraphics[height=5.8cm]{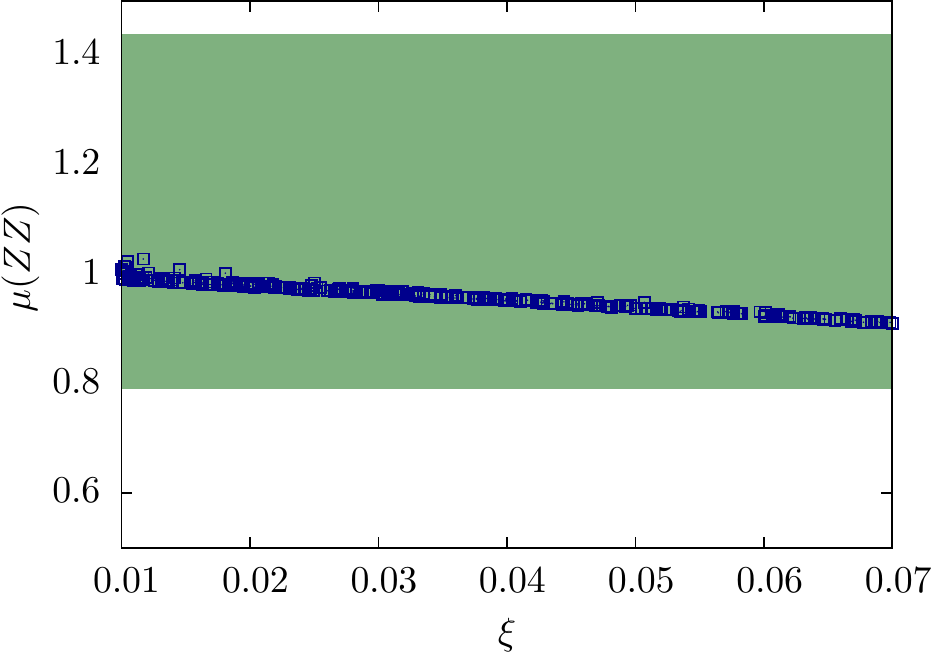}}\\
  \subfigure[~$WW$ signal
  strength.]{\includegraphics[height=5.8cm]{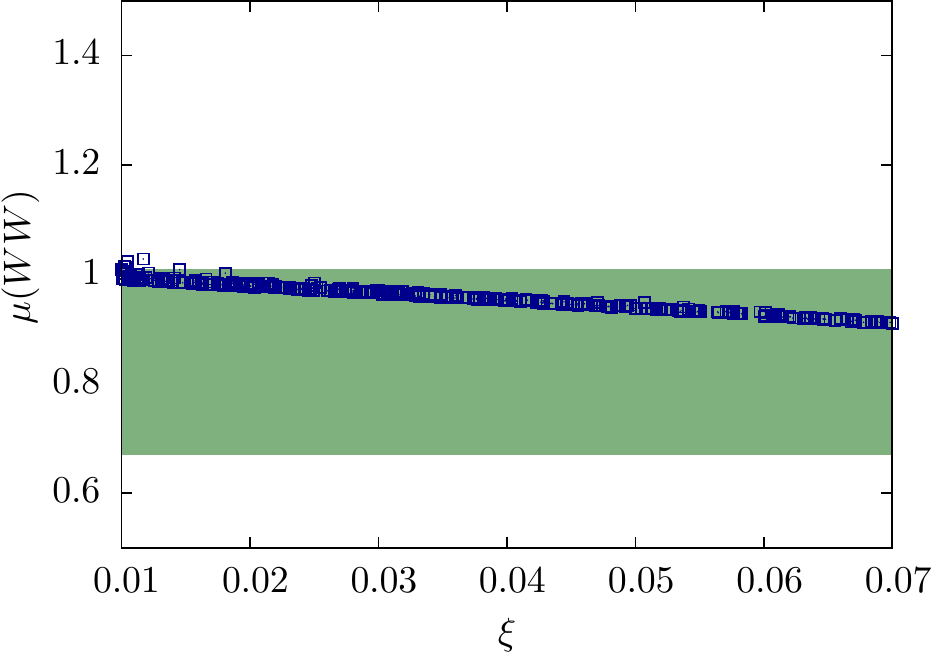}} \hfill
  \subfigure[~$\tau\tau$ signal
  strength.]{\includegraphics[height=5.8cm]{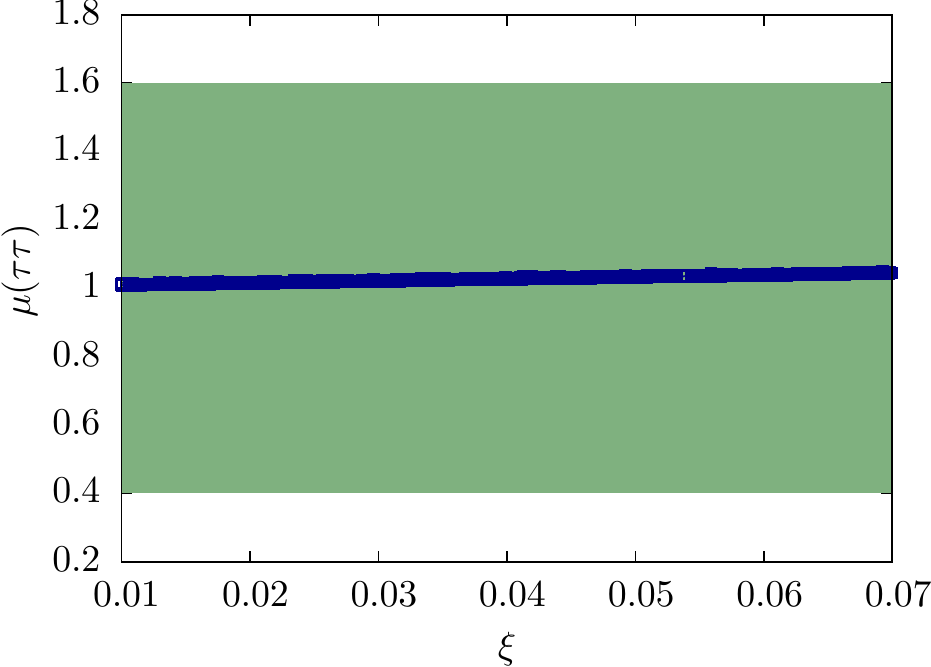}}
  \caption{\label{fig:scan} 125 GeV Higgs signal strengths as
    constrained by the ATLAS and CMS combination of
    Ref.~\cite{Khachatryan:2016vau}. The blue-shaded area corresponds
    to the points in our scan yielding the correct top and bottom
    masses. The red points in panel (a) result from a modified scan
    which includes the charged exotic Higgs loop contributions to the
    diphoton partial decay width, demonstrating that the signal
    strength in the photon channel can be significantly impacted by
    the presence of these states. The scatter in the red points results from varying the sign and size of the unknown trilinear Higgs couplings.}
\end{figure*}

Searches for pair-produced colour-octet scalars $\pi^a$, as predicted
from the breaking to QCD in Eq.~\eqref{eq:symbreak}
$\SU(3) \times \SU(3)' \to \SU(3)_c$ with subsequent gauging of QCD, have
been considered in theories of vector-like
confinement~\cite{Kilic:2009mi,Kilic:2010et,Schumann:2011ji},
compositeness~\cite{Hayot:1980gg,Ellis:1980hz,Belyaev:1999xe,Bai:2010mn,Ferretti:2016upr},
as well as in hybrid SUSY
models~\cite{Plehn:2008ae,Choi:2008ub}. Searches were performed during
Run-1~\cite{ATLAS:2012ds,Khachatryan:2014lpa} in four jet final states
as well as in R-parity violating SUSY
scenarios~\cite{Aad:2016kww}. CMS have published a search using first
13 TeV data which pushes constraints into the multi-TeV
regime~\cite{Khachatryan:2015dcf}. None of the analyses have reported
anomalies or even evidence; the $\pi^a$ color octet mass is therefore
pushed to $m_{\pi^a} \gtrsim 3.5~\text{TeV}$ by using the results
of~\cite{Khachatryan:2015dcf} (assuming $\text{BR}=1$).  While this
scale is important information for non-perturbative analyses (e.g.
\cite{Knecht}), it does not impact the model's phenomenology in the
weak sector. A comprehensive analysis of the phenomenology of these
states was presented recently in Ref.~\cite{Belyaev:2016ftv}.

\subsection*{Constraints from Higgs signal strengths}
As already mentioned, the phenomenology of the 125 GeV SM-like Higgs
boson follows largely the MCHM4/5 paradigm, with one crucial difference
related to the potential appearance of additional charged exotic Higgs
bosons which could modify the Higgs signal strengths, which we define
as
\begin{equation}
	\mu= {\sigma \text{BR} \over [\sigma \text{BR}]^{\text{SM}}}\,.
\end{equation}
$\sigma$ and $\text{BR}$ denote the production cross section and
branching ratios for $gg \to \Phi \to ( WW,ZZ,\gamma\gamma,\tau\tau)$ respectively. 
We will limit ourselves to the
dominant gluon fusion production mode in this work.  The signal
strengths are relatively precisely determined quantities after
Run-1~\cite{Khachatryan:2016vau} (see also~\cite{Aad:2015pla} for an
interpretation of ATLAS results in terms of composite models).

In Fig.~ \ref{fig:scan}, we show a scan over the model following the
prescription as detailed earlier. As can be seen, the current Higgs
signal strength measurements are consistent with the model's
prediction over a large range of values of $\xi=v^2/f^2$. In this
sense our findings are consistent with the analysis
of~\cite{Aad:2015pla}. However, the possibility of additional charged
scalars running in the $h\to \gamma\gamma$ loops can significantly
change this result\footnote{Similar ideas have been used to explain
  the early excess in the observed diphoton branching ratio,
  see~\cite{Akeroyd:2012ms}.}. Given the early stage of the Higgs
phenomenology programme, the Higgs measurements are not sensitive
enough to provide tight constraints on the model.

\subsection*{Constraints from exotic Higgs searches}

\subsubsection*{Doubly Charged Scalars}
The most striking BSM signature related to the exotic Higgs states is
the production of doubly charged scalars. Since the triplet states'
potential is not affected by electroweak symmetry breaking, these
states can only be pair-produced as $W^+W^+\phi^{-}_{-}$ vertices are absent
in the effective theory. This leads to a qualitatively different
phenomenology compared to one of the standard scenarios of scalar weak
triplets~\cite{Georgi:1985nv,Chanowitz:1985ug,Gunion:1989ci}: In our
case, the dominant production mechanism relevant for the LHC is
Drell-Yan production (with expected moderate QCD corrections
$K\simeq 1.3$~see e.g.~\cite{Hamberg:1990np}) which is entirely
determined by the hypercharge and $\SU(2)_L$ quantum numbers of the
doubly charged scalar. For a choice $m_{\phi^{\pm}_{\pm}}=200~\text{GeV}$,
we obtain a Drell-Yan cross section of $84~\text{fb}$\footnote{We use
  a combination of
  {\sc{Feynrules}}~\cite{Christensen:2008py,Alloul:2013bka,Alloul:2013jea},
  {\sc{Ufo}}~\cite{Degrande:2011ua} and
  {\sc{MadEvent}}~\cite{Alwall:2014hca} for the calculation of the
  cross section.}, which decreases exponentially for heavier masses.

Current analyses~\cite{ATLAS:2014kca,Chatrchyan:2012ya} set
constraints mostly from searches for same-sign lepton production,
which are motivated from a Majorana-type lepton sector operators
involving the ${\bf{3}}_1$ multiplet in the Georgi-Machacek
model~\cite{Georgi:1985nv,Gunion:1989ci}. Although leptons are not
included in  Ferretti's proposal~\cite{Ferretti:2014qta}, we can
expect the biggest coupling to arise from $\tau$ leptons following the
partial compositeness paradigm. Ref.~\cite{Chatrchyan:2012ya} sets a
constraint in this channel of $\sim 100~\text{fb}$, which is not
stringent enough to constrain the presence of a doubly charged Higgs
boson as predicted in the model even when we consider decays to $\tau$
leptons.

If this lepton operator is not considered, the dominant decay will be
to same sign $W$ bosons via fermion
loops~\cite{Englert:2016ktc}. Ref.~\cite{ATLAS:2014kca} does not make
any specific assumptions on jet or missing energy activity and set
constraints of $\sim 1~\text{fb}$. Including the $W$ branching
fractions the weak pair production of the doubly-charged scalar in our
model readily evades these constraints. The recent
analysis~\cite{Englert:2016ktc} that specifically targets the
$pp\to 4\ell + \text{missing energy}$ smoking signature shows that the
LHC should in principle be able to probe a mass regime up to 700~GeV.

\subsubsection*{Charged Scalars}

Charged Higgs boson searches have been performed during Run-1 by
ATLAS~\cite{Aad:2015typ} and CMS~\cite{Khachatryan:2015qxa} from the
production off top quarks and set constraints of $0.6$-$0.8~\text{pb}$
in the considered mass region. In our scan, we find cross
sections\footnote{Again we use a combination of
  {\sc{Feynrules}}~\cite{Christensen:2008py,Alloul:2013bka,Alloul:2013jea},
  {\sc{Ufo}}~\cite{Degrande:2011ua} and
  {\sc{MadEvent}}~\cite{Alwall:2014hca}} in the range of
$\simeq 1~\text{fb}$ after averaging between the 4 and 5 flavour scheme
as detailed in~\cite{Harlander:2011aa}.  W conclude that available
LHC analyses are not sensitive enough to constrain the exotic Higgs
spectrum because of the small production cross section.

\subsubsection*{Neutral Scalars}
The interactions of Eq.~\eqref{eq:massint} also introduces Yukawa-type
interactions with the heavy SM fermions and top partners after
diagonalisation of Eqs.~\eqref{eq:topmass} and
\eqref{eq:bottmass}. The dominant production modes of the extra
neutral scalars is then gluon fusion with heavy SM fermions and top
partners running in the gluon fusion loops.\footnote{There is also the
possibility of small anomaly-induced terms which we will not consider
in this work; they are expected to be parametrically small \cite{Ferretti:2016upr}.}

\begin{figure}[t]
  \includegraphics[height=5.8cm]{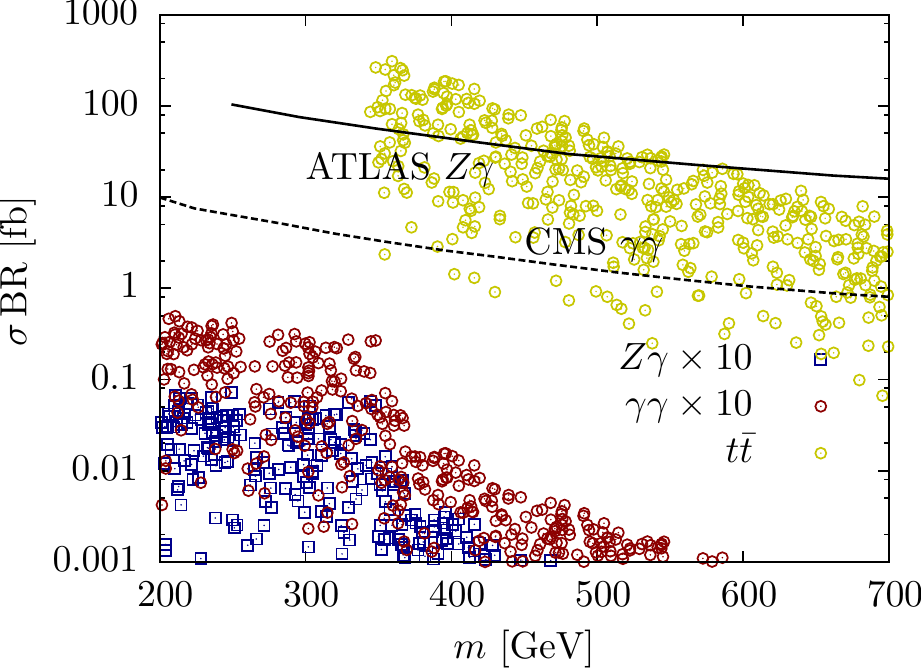}
  \caption{\label{fig:31bosons1} Scan over the neutral, CP even
    ${\bf{3}}_1$ state including ATLAS~\cite{ATLAS-CONF-2016-044} and
    CMS~\cite{Khachatryan:2015qba,Khachatryan:2016yec}. Currently no model-independent LHC constraint exists for the $\bar t t$-channel.}
\end{figure}

\begin{figure}[b]
  \includegraphics[height=5.8cm]{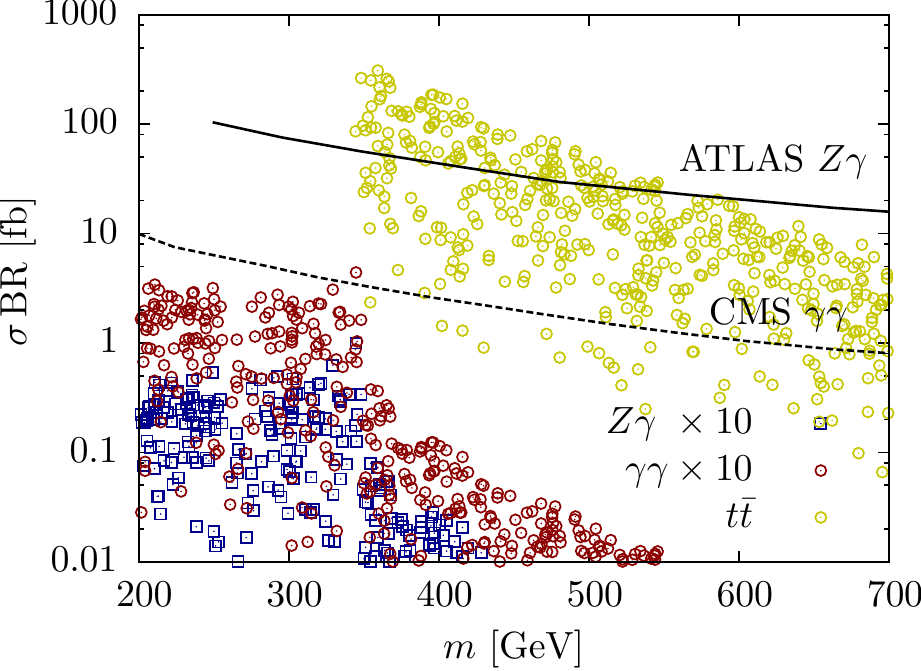}
  \caption{\label{fig:31bosons2} Scan over the neutral, CP odd
    ${\bf{3}}_1$ state including ATLAS~\cite{ATLAS-CONF-2016-044} and
    CMS~\cite{Khachatryan:2015qba,Khachatryan:2016yec}. Currently no model-independent LHC constraint exists for the $\bar t t$-channel.}
\end{figure}

\begin{figure*}[!t]
  \centering
  \subfigure[~Mass of the lightest top partner.]{\includegraphics[height=5.8cm]{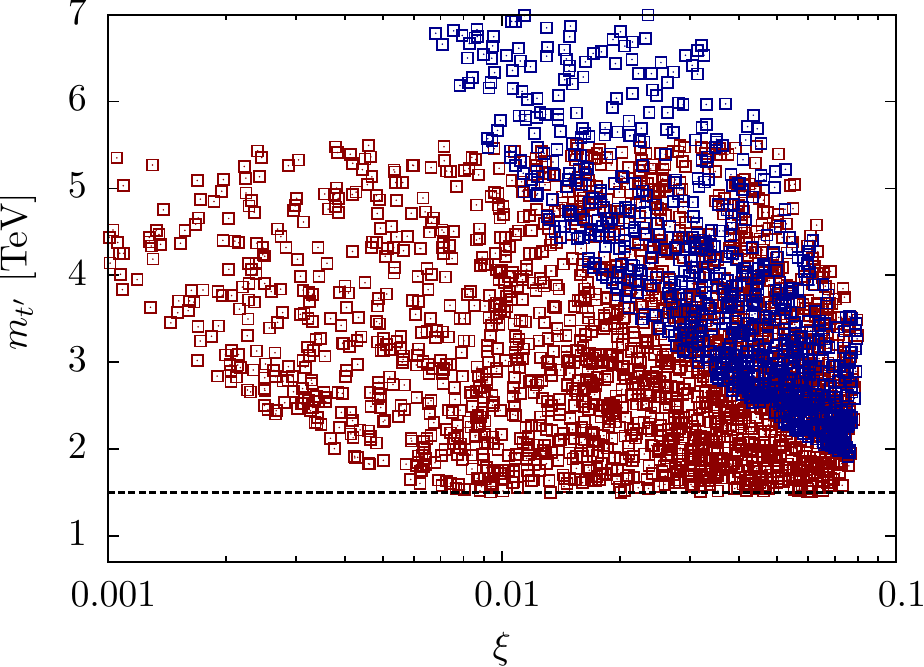}}\hfill
  \subfigure[~Diphoton signal strength.]{\includegraphics[height=5.8cm]{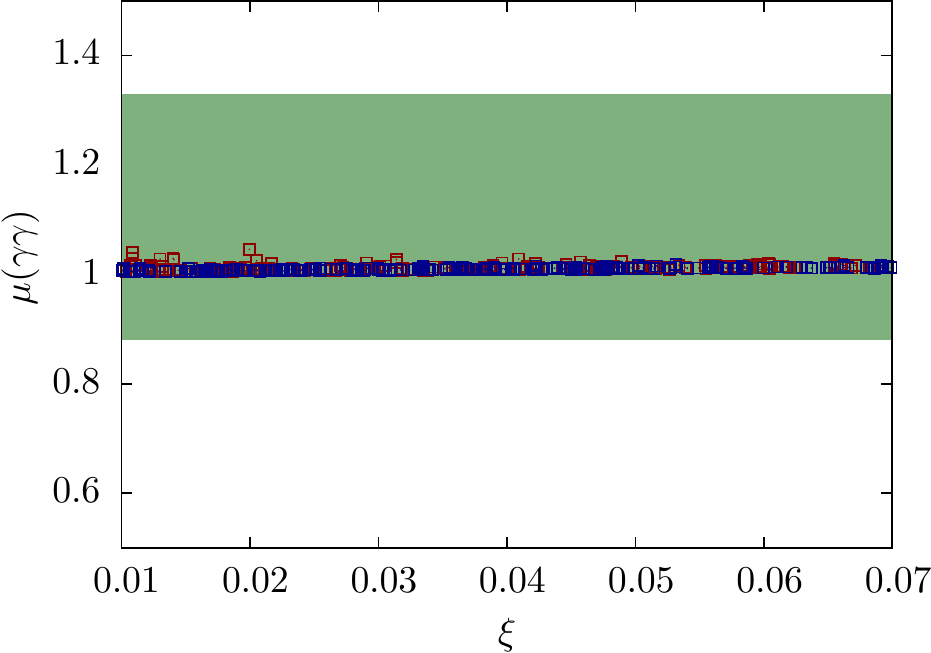}}
  \caption{Contour plot for a scan of the first non-SM top partner in
    agreement with the current LHC constraints (dashed line) detailed in
    Sec.~\ref{sec:results} (a) and diphoton Higgs signal strength (b). Blue points show the correlation expected
    from a lattice result of $ \hat{M} \in[2,5]$ while the red
    points leave $M$ as a free parameter in
    $M\in [1.5,5.5]~\text{TeV}$.}
  \label{fig:toppartners}
\end{figure*}

We calculate
the gluon fusion cross sections,\footnote{Using  a modified version of
{\sc{Vbfnlo}}~\cite{Arnold:2008rz} together with
{\sc{FeynArts}}/{\sc{FormCalc}}/{\sc{LoopTools}}~\cite{Hahn:1998yk,Hahn:2000kx}.}
for the parameters that reproduce the correct top and bottom masses,
which satisfy  constraints of the  current top partners outlined above as well as  
the 125 GeV Higgs measurements. 
A flat QCD $K\simeq 1.6$
factor~\cite{Dawson:1990zj,Kauffman:1993nv,Graudenz:1992pv,Spira:1995mt,Spira:1995rr} is included.

Since the ${\bf{3}}_0$ state couples to
$\sim \lambda_q \bar b_L B_R /\sqrt{2}+{\text{h.c.}}$ the
phase space enhanced decay into physical bottom quarks dominates, irrespective of
the smallness of the coupling. For these final states there are
currently no sensitive searches given the large expected QCD
backgrounds and the challenge of triggering such final states in the
first place.

Loop-induced decays (see Appendix~\ref{app:loop}) to $\gamma \gamma$
are already fairly constrained after Run-1. For instance, CMS limit
$\sigma{\text{Br}}(\gamma\gamma)\lesssim 1$-$10~\text{fb}$ between
180 and 800 GeV with little dependence on the resonance
width~\cite{Khachatryan:2015qba} (see also the analysis by
ATLAS~\cite{Aad:2014ioa} with similar sensitivity). CMS have updated
their results also including 13 TeV data~\cite{Khachatryan:2016yec},
which mostly extends the sensitivity region up to
$m\simeq 4~\text{TeV}$ with limits
$\sigma{\text{Br}}(\gamma\gamma) \lesssim 0.2~\text{fb}$ for
$m>2~\text{TeV}$. Numerically we find the diphoton branching ratios to
be suppressed by three orders of magnitude compared to $b\bar b$ for
the ${\bf{3}}_0$ state in our scan, which leaves it unconstrained by
these measurements (identical conclusions hold for other loop-induced
decays).

The neutral ${\bf{3}}_1$ states do not couple to bottom quarks but
both CP-even and odd interactions follow from the operator
$\sim \sqrt{2}\lambda_q\, \bar t_L Y_r + {\text{h.c}}$. This opens up
the interesting phenomenological possibilities below the $t\bar t$
threshold. We find that for such a mass choice the decay into gluons
typically dominates.\footnote{We retain full mass dependencies and
  include all non-diagonal Higgs interactions in the decay diagrams at
  one-loop. We consider decays to $ZZ$,~$WW$,~$gg$,~$\gamma \gamma$
  and $Z\gamma$.} However, it is worthwhile to also check the
sensitivity to these states in other final states, also extending
beyond the aforementioned diphoton analysis.

The production of $Z\gamma$ final states was constrained in Run-1
analyses~\cite{Chatrchyan:2013vaa,Aad:2014fia}, which focused on mass
ranges inspired by the SM $m\lesssim 190~\text{GeV}$ with only weak
constraints $\sigma \text{BR}(Z\gamma) \lesssim 100~\text{fb}$. ATLAS
and CMS have extended these searches to the higher mass
regime~\cite{ATLAS-CONF-2016-044,CMS-PAS-EXO-16-034} with 13 TeV data
and set limits ${\cal{O}}(10)~\text{fb}$ above 300 GeV. The hierarchy
in branching ratios, however, makes neither the diphoton searches nor
the $Z\gamma$ analyses sensitive enough to impose mass limits on the
considered CP even state, Fig.~\ref{fig:31bosons1}.

Searches for $ZZ$ and $WW$ decays, which are also
mediated at the loop level are available~\cite{Khachatryan:2015cwa,ATLAS-CONF-2016-082}
and constrain signal strengths of $\sim 10\%$ relative to the SM
expectation. These
searches are not yet sensitive enough to constrain this scenario.

A similar conclusion holds for the CP odd state as the increase in
production cross sections is not sizable enough to make current
constraints sensitive to the model. The bulk of the considered
parameter space is left constraint with the early 13 TeV data,
Fig.~\ref{fig:31bosons2}. Once the $t\bar t$ channel becomes
accessible as a decay mode, the loop-induced decays become
unconstrained for scalar masses above $2m_{t}$.

If the mass of the neutral scalar lies above the top mass threshold,
the decay to top pairs becomes kinematically accessible and will
dominate over the loop-induced diboson decays. Searches for the CP
even or odd scalar resonances in $t\bar t$ final states exist in the
context of two Higgs doublet
models~\cite{ATLAS-CONF-2016-073}. Although this analysis is difficult
to interpret in our scenario due to the involved signal-background
interference, the sensitivity in this search probes
$\tan\beta\simeq 1$, which corresponds to a signal cross section of
around 0.15~pb around 500 GeV which quickly decreases for larger
masses. As can be seen from Figs.~\ref{fig:31bosons1} and
\ref{fig:31bosons2}, this search will start to constrain the parameter
space, although the spread of points shows that there is still a large
range parameter points where the model remains viable, in particular
for larger masses.

Ignoring systematic uncertainties in extrapolating the results to the
high luminosity target of 3/ab, the CMS $\gamma\gamma$ analysis should
significantly constrain the presence of extra scalars in the spectrum
below the $t \bar t$ threshold as the exclusion contour will be become
a factor $\sim 15$ more stringent. A similar conclusion holds for the
$t\bar t$ channel although details will depend on signal background
interference.

In summary, we find that while there are searches at the LHC which
might become sensitive to the exotic states predicted by the model of
Sec.~\ref{sec:model} in the near future, current analyses are not yet
constraining enough to significantly limit the models parameter
space. This can be understood as a motivation to explore this scenario
on the lattice as valid candidate theory of TeV scale compositeness.

Finally, coming back to the potential impact of lattice input, we show
the scan of top partners assuming a lattice calculation input of
$\hat{M}$. This results in a correlation of the top partner spectrum
with $f$, Fig.~\ref{fig:toppartners} and indicates that an observation
of top partners in the near future at the LHC can not only provide an
input to a more comprehensive investigation on the lattice, but, more
importantly can potentially rule out the model of Eqs.~\eqref{eq:break}
and~\eqref{eq:symbreak} directly.

\section{Summary and Conclusions}
\label{sec:summary}
The observation of a SM-like Higgs and no additional evidence of
physics beyond the SM provides no hint towards a more fundamental
theory of the TeV scale.

Non-minimal theories of Higgs compositeness have always been
attractive solutions to solve this puzzle, but recently they have
received particular attention as the possibility of UV-complete models
paves the way for applying non-perturbative techniques. Such a
programme needs to be informed by the results of the LHC as collider
constraints can be understood in terms of the UV-theory's LECs. In
this work, we provide the latest constraints from Higgs-like
measurements as well as from searches for additional pseudo
Nambu-Goldstone weak triplets with exotic charges predicted by the
scenario of~\cite{Ferretti:2014qta}.

Including constraints from the literature on the exotic states that
are relevant for our analysis of LECs of this particular scenario, we
find that the latter is largely unconstrained at this stage in the LHC
programme. Extrapolating to $3/ \text{ab}$, the weak exotics searches
are capable of limiting the effective theory's parameter space. In
particular, the increasing precision on the 125 GeV Higgs couplings
(see e.g.~\cite{Bock:2010nz,Klute:2013cx,Englert:2014uua}) will allow
us to explore the coupling strength deviations in the $ 5\%$-range,
which will provide stringent constraints (see Fig.~\ref{fig:scan}) on
the model.

Direct searches are not constraining on the top partner mass $m_t'$
but when combined with lattice determinations the situation may
change.  For instance, the prediction of the hypercolor baryon mass
$M$, in units of the decay constants $f$, provides directly
falsifiable predictions on the top quark partner spectra as shown
in~Fig.~\ref{fig:toppartners}.  In the longer term, the computation of
the Higgs potential parameters $\al$ and $\be$ provides
first principle constraints on the viability of the model against the
Higgs mass and Higgs decay channel measurements
(cf. Fig.~\ref{fig:ABPlot}).  In particular the determination of only
one of these parameters can exclude the model whereas both parameters
are needed to confirm it in this sector.  The lattice technology
developed within this particular model can be used for future UV
completions that may become interesting in the future.

\bigskip
\noindent {\emph{{Acknowledgments}}} --- We thank Tom DeGrand,
Gabriele Ferretti, Marc Gillioz, Maarten Golterman, and Yigal Shamir
for helpful discussions and especially Gabriele Ferretti for relevant
comments on the draft.  In particular we would like to thank Maarten Golterman, and Yigal Shamir for informing us about the possibility of further 
$4$-point function corrections to the Higgs potential.
LDD and RZ are supported by the
UK STFC grants ST/L000458/1 and ST/P000630/1. LDD is supported
by the Royal Society, Wolfson Research Merit Award, grant WM140078.

\appendix

\section{Four point-functions}
\label{app:4}
The Higgs potential arises from integrating out the gauge bosons using the Coleman-Weinberg method (giving rise to $\CLR$) 
and involves the effective top and bottom quark couplings to the hypercolour-baryon. The 
contributions of the latter are given by 4-point correlator functions
\begin{alignat}{3}
&  \CtLR  &=& - (\la_1 \la_2)^2  & & \int_{x_{1,2,3}} \vev{ J_{\bar{L}Ri}(x_1,x_2)  
J^\dagger_{\bar{L}Rk}(x_3,x_4)    }_{i \neq k}  \;, \nonumber \\
&  \CRRRR   &=& - \la_2^4 \times  & & \int_{x_{1,2,3}} \vev{J_{RRi}(x_1,x_2)  
J^\dagger_{RRk}(x_3,x_4)    }_{i \neq k} \;, \nonumber \\
\label{eq:appr1}
&   \CLLLL  &=& - \la_1^4 \times & & \int_{x_{1,2,3}} \vev{J_{LLi}(x_1,x_2)  J^\dagger_{LLk}(x_3,x_4)    }_{i \neq k} \;,
\end{alignat}
where $\CtLR = - y^2 \Ctop$ in the notation of Ref.~\cite{Golterman:2015zwa}, and where we have defined the short-hand notation
$\int_{x_{1,2,3}} = \int d^4 x_1\,d^4 x_2\,d^4 x_3$. The bi-local currents in Eq.~\eqref{eq:appr1} are defined as
\begin{alignat*}{2}
& J_{\bar{L}Ri}(x_1,x_2)  &\; =\; &  \bar{t}_L \By_i(x_1) \bBy_i t_R(x_2) \;, \nonumber  \\
& J_{RRi}(x_1,x_2)  &\; = \; &   \bBy_i(x_1) t_R \bBy_i t_R(x_2) \;,  \\
& J_{LLi}(x_1,x_2)  &\; =\; &   \bBy_i(x_1) t_L \bBy_i t_L(x_2) \;. \nonumber
\end{alignat*}
The latter originate from the (E)HC interaction 
\begin{equation*}
\label{eq:EHC}
{\cal L}_{\textrm{EHC}} = \la_1 \bar{\hat{q}}_{L}  \By_R + \la_2 \bar{\hat{q}}_{R}  \By_L + \textrm{h.c.} \;,
\end{equation*}
with $\hat{q}_{L,R} = T_{L,R}$  in the notation of Ref.~\cite{Golterman:2015zwa}. The
hypercolour-baryon operator is given by
\begin{equation}
\By_{Ria} = - \frac{1}{2}  \eps^{ABCD} \eps_{abc} P_R \psi_{AB i } \chi^T_{Cb} C P_R \chi_{Dc} \;,
\end{equation}
where $a,b,c$ are $\SU(3)_c$, $A,B,C,D$ are $\SU(4)_{\textrm{HC}}$, and $i$ is a
$\SO(5)$ indices.
Comparing to the notation of \cite{Golterman:2015zwa}, we use $\chi \leftrightarrow 
\psi$ in accordance with Ferretti's original convention. 
Note that ${\hat{q}}_L$ but not $\hat{q}_R$ transforms non-trivially under 
the effective custodial symmetry $SU(2)_R$. 
Therefore $\la_1$ or $\CLLLL$ 
are responsible for further splittings of the two isotriplets 
$SU(2)_L$ in Eq.~\eqref{eq:split}.

At last we note that in order to obtain a potential which is manifestly $\SU(2)_L$ invariant 
the bottom quarks also needs to be integrated out.  The $4$-point function, focusing 
on the top quark, do though give the right coefficients.

\section{Analysis of loop-induced decays of the non-Higgs scalar states}
\label{app:loop}
In this section we briefly review the calculation underpinning the
loop-induced decays of the additional neutral scalars in the model.

\begin{figure}[!b]
  \includegraphics[width=4.2cm]{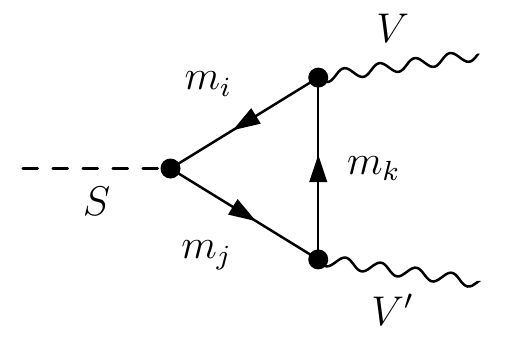}
  \caption{\label{fig:dec} Representative Feynman diagram mediating
    the decay of a neutral scalar $S\in\{{\bf{3}}_0,{\bf{3}}_1\}$ to
    vector bosons $V,V'\in\{Z,\gamma,W^\pm\}$ with interaction
    vertices obtained in the mass-diagonal representation of the
    charged and neutral top and bottom space currents.}
\end{figure}

After diagonalising the top- and bottom mass mixing matrices, the
scalar as well as vectorial couplings will be in general non-diagonal
in the top and bottom partner spaces (and not necessarily purely vectorial)w . This leads to a multi-scale
decay amplitude that can be pictorially represented by the sum over
Feynman diagrams indicated in Fig.~\ref{fig:dec}.

We can write the unrenormalised decay amplitude at one loop as
\begin{equation}
  \label{eq:dec}
  i{\cal{A}} = \sum_i C_i \langle \hat O_i \rangle 
\end{equation}
with $\hat{O}_i$ denoting the quantum operators contributing to the
decay with matrix element $\langle \hat O_i \rangle$ and associated
couplings $C_i$ (which can have a non-zero mass dimension). In our
case the relevant operators are
\begin{alignat*}{5}
  \hat O_1 &= \hat S \hat V^\mu \hat V'_\mu \,,\\
  \hat O_2 &= \hat S \hat V^{\mu\nu} \hat V'_{\mu\nu} \,,\\
  \hat O_3 &= \hat S \hat{V}^{\mu\nu} {\hat{\widetilde{
    V'}}}_{\mu\nu}\,,
\end{alignat*}
$\tilde V$ denotes the dual field strength tensor.

The latter two operators typically arise from integrating out chiral
fermions~\cite{Kniehl:1995tn}, while the first one is the standard
$V$-Higgs interaction associated with spontaneous symmetry breaking.

Calculating the decay amplitude, one finds that all coefficients in
Eq.~\eqref{eq:dec} are UV-finite except $C_1$. This result is familiar
from the SM within which top and bottom loop contributions renormalise
the tree-level $HVV$ operators.

There is no such interaction in the EFT for the non-Higgs states and
the considered order in chiral perturbation theory due to the symmetry
of the underlying UV theory. However, these symmetries are spurious
(e.g.~leading to $S$ obtaining a mass from a Coleman-Weinberg
potential) and quantum corrections will excite all operators that are
allowed by explicitly intact symmetries. Hence, they will also excite
the absent operator $\hat{O}_1$.

It is interesting to mention another similarity with the $\SU(2)_L$
triplet scenario of~\cite{Georgi:1985nv} here. In this model, the
requirement of custodial invariance identifies the real and complex
$\SU(2)_L$ triplet vacuum expectation value. This identification is
broken at the quantum level signalised by the appearance of additional
UV singularities that require the introduction of independent bare
quantities~\cite{Gunion:1990dt} (while the renormalised quantities can
be identified as input to the renormalisation procedure).

From a technical perspective the problem encountered in the
calculation of the decay amplitude is similar. We are forced to
introduce a bare operator $\hat{O}_1$ and supply the underlying UV
physics through a renormalisation condition (as part of a minimal set
of EFT input parameters). The counter term amplitude in our case can be
written as (using $C_1=C_1^R+\delta Z_{C_1}$ and multiplicative
renormalisation of the quantum fields) turns Eq.~\eqref{eq:dec} into
using
\begin{equation}
  i{\cal{A}} = (C_1 + \delta Z_C) \langle \hat O_1 \rangle^R + \dots
\end{equation}
with the ellipses denoting finite terms
$\sim \langle \hat O_{1,2} \rangle$.

To reflect the symmetry breaking pattern that underpins our EFT
formulation, we need to provide input data to the renormalisation
procedure. This is guided by the EFT not allowing the dimension three
operator due to the approximate shift symmetry of the Nambu-Goldstone
$S$. All interactions generated by loops that violate this symmetry
(and eventually creates a mass of $S$) are higher order in the EFT
expansion~\cite{Giudice:2007fh,Kniehl:1995tn}. A suitable
renormalisation condition is therefore a vanishing coefficient
$C^R_1=0$. This fixes the renormalisation constant $\delta Z_{C_1}$
and renders the amplitude UV finite.


\bibliography{paper.bbl}

\end{document}